%% file: main.tex
\documentclass[conference]{IEEEtran}
\IEEEoverridecommandlockouts
\usepackage{amsmath,amsfonts}
\usepackage{algorithmic}
\usepackage{algorithm}
\usepackage{array}
\usepackage[caption=false,font=normalsize,labelfont=sf,textfont=sf]{subfig}
\usepackage{textcomp}
\usepackage{stfloats}
\usepackage{url}
\usepackage{verbatim}
\usepackage{graphicx}
\usepackage{cite}
\usepackage{csquotes}
\usepackage{amssymb}
\usepackage{nopageno}

\usepackage{tikz,stackengine,mathtools}
\usetikzlibrary{matrix,calc}
\newsavebox{\foobox}

\usepackage{tikz}
\usetikzlibrary{arrows,shapes.misc,chains,scopes}
\usepackage{pgfplotstable}
\pgfplotsset{compat=1.15}

\title{Turbo product decoding of cubic tensor codes 
}
\begin{document}

\author{\IEEEauthorblockN{Sarah Khalifeh}
\IEEEauthorblockA{\textit{Dept. ECE} \\
\textit{Northeastern University}\\
Boston, USA \\
khalifeh.s@northeastern.edu}
\and
\IEEEauthorblockN{Ken R. Duffy}
\IEEEauthorblockA{\textit{Dept. ECE \& Math} \\
\textit{Northeastern University}\\
Boston, USA \\
k.duffy@northeastern.edu}
\and
\IEEEauthorblockN{Muriel M{\'e}dard}
\IEEEauthorblockA{\textit{Research Laboratory for Electronics} \\
\textit{Massachusetts Institute of Technology}\\
Cambridge, USA \\
medard@mit.edu}

}

\maketitle

\begin{abstract}
Long, powerful soft detection forward error correction codes are typically constructed by concatenation of shorter component codes that are decoded through iterative Soft-Input Soft-Output (SISO) procedures. The current gold-standard is Low Density Parity Check (LDPC) codes, which are built from weak single parity check component codes that are capable of producing accurate SO. Due to the recent development of SISO decoders that produce highly accurate SO with codes that have multiple redundant bits, square product code constructions that can avail of more powerful component codes have been shown to be competitive with the LDPC codes in the 5G New Radio standard in terms of decoding performance while requiring fewer iterations to converge. Motivated by applications that require more powerful low-rate codes, in the present paper we explore the possibility of extending this design space by considering the construction and decoding of cubic tensor codes. 
\end{abstract}


\thispagestyle{empty}

\section{Introduction}
Since the earliest days of forward error correction it has been known that an effective way to construct long, powerful codes that can be practically decoded is to concatenate short component codes, \cite{elias_error-free_1954,forney1966_concatenated,costello2007channel}. In the presence of soft-input, such constructions can be efficiently decoded iteratively with high levels of latency-reducing parallelism using a Soft Input Soft Output (SISO) decoder for each component code. Schemes of this sort include 
low-density parity-check (LDPC) codes \cite{gallager1962ldpc, mackay1997near, richardson2001efficient,mansour2003high,hailes2015survey} and turbo product codes (TPCs) \cite{pyndiah_1998}.

Central to the performance of any SI iteratively decoded code is the structure through which the code is constructed from components, and the quality of the SO of the component code decoder. LDPC codes, for example, are constructed with the weakest possible component code, single parity checks, from which essentially perfect SO can be computed. In contrast, TPCs 
can be built with more powerful component codes, such as extended Bose-Chaudhuri-Hocquenghem (eBCH) codes, but
the original TPC decoder proposed in \cite{pyndiah_1998} was restricted to using component codes that had an efficient hard detection decoder from with approximate bit-wise SO could be produced from a list of possible decodings. Certain modern SI decoders can provide a list decoding for a broader class of component codes, including Polar codes \cite{niu2012crc,tal2015_list,balatsoukas2015llr} and arbitrary moderate redundancy codes \cite{Duffy19a,duffy22ORBGRAND, abbas2021list}, which greatly expands the class of component codes and code dimensions from which product codes can be decoded with Pyndiah's approximate SO leading to further exploration of their capabilities \cite{condo2022_iterative,galligan2023_block,cocskun2024precoded}.

Recently, it has been established that SI Guessing Random Additive Noise Decoding (GRAND) algorithms, which can decode any moderate redundancy component code of any structure, can readily provide substantially more accurate SO than Pyndiah's approximation, and that its use in iterative decoding of product codes improves decoding performance \cite{yuan2023soft}. By considering Guessing Codeword Decoding (GCD) \cite{ma2024guessing} through the lens of GRAND, it has been established that it too can generate accurate SO \cite{duffy2024SOGCD}, as can Successive Cancellation decoding of Polar-like codes \cite{yuan2024nearoptimal,yuan_24}. When TPCs are decoded with these decoders that provide improved SO, their block error rate (BLER) and bit error rate (BER) are sufficiently enhanced that they can outperform the LDPC codes used in the 3GPP 5G New Radio (NR)~\cite{richardson2018design} standard \cite{yuan2023soft,duffy2024SOGCD,yuan2024nearoptimal}. Half-decoding iterations form the minimal parallelizable unit in the decoding process and these TPCs also require fewer iterations to complete their decoding than an LDPC.

Traditionally, wireless communication applications required codes with rates of $\approx0.2$--$0.95$ and lengths of the order of $100$s--$1000$s of bits to meet system demands, resulting in LDPC codes being selected for, e.g., the 5G New Radio data channel. For communications in challenged environments \cite{lanzagorta2012underwater} as well as emerging applications, such as continuous key quantum key distribution (CV-QKD)~\cite{bennett1992experimental}, however, there is a need for more powerful, lower rate codes that are still efficiently decodable. 

\begin{figure}[h]
\begin{center}
\includegraphics[width=1\columnwidth]{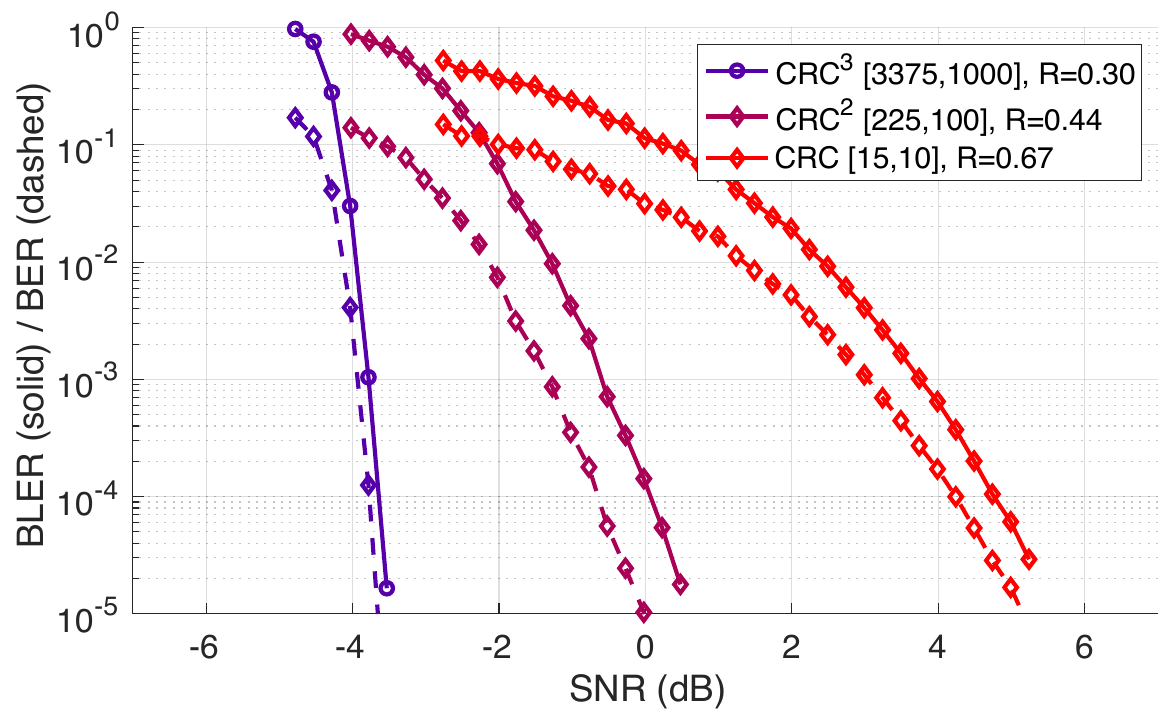}
\end{center}
\caption{Block error rate (solid) and bit error rate (dashed) vs SNR of a $[15,10]$ component, a $[225,100]$ product, and a $[3375,1000]$ cubic code constructed with a CRC component code with polynomial 0x15 in Koopman notation \cite{koopman2009cyclic}. Component code decoding is performed with 1-line ORBGRAND and iteratively decoded with a SISO version for the product and cubic code.} 
\label{fig:15_10_cubic}
\end{figure}

Leveraging those recent developments of accurate SISO decoders that can decode powerful component codes, towards meeting new application demands here we explore the performance of a natural generalization of iterative decoding of TPCs that enables the construction of longer, lower-rate codes: cubic tensor product codes. If a component code takes $k$ information bits to $n$ codeword bits for a code-rate of $R=k/n$, a standard product code has dimensions $(n^2,k^2)$ for a rate $R^2$. A cubic tensor product code, however, has dimensions
$(n^3,k^3)$ giving a rate of $R^3$, which is a significantly longer, lower-rate code. Crucially, cubic tensor codes can still be decoded with extremely high levels of parallelizability that would result in desirable low-latency decoding when implemented in circuits. As with square product codes, we shall see that cubic tensor codes also complete their decoding in a small number of iterations as a consequence of having multiple parity checks per component code.

An illustration of the code dimensions that are readily available with this construction as well as block error rate (BLER) and bit error rate (BER) performance from such codes is provided in Fig. \ref{fig:15_10_cubic}. Starting with a $[15,10]$ cyclic redundancy check (CRC) component code that takes $k=10$ information bits to produce an $n=15$ bit codeword, a $[15^2,10^2]=[225,100] $ product code and a $[15^3,10^3]=[3375,1000] $ cubic code are constructed. All three codes are decoded with 1-line ORBGRAND \cite{duffy22ORBGRAND}, with the product code and cubic code decoded with the SISO version \cite{yuan2023soft}. While more detail will be provided in the following sections on the construction and decoding of these codes, it can be seen that the cubic tensor code provides a much longer, more powerful code suitable for more challenged environments.

The rest of this paper is structured as follows. In Section \ref{sec:tensor}, we explain the tensor code construction and the parallelizability of its encoding. In Section \ref{sec:TPC}, we explain the decoding algorithm, which is built on having accurate SISO component decoders. While Pyndiah's original update rule for TPC decoding had two hyperparameters per half-iteration, use of the recent SISO decoders removes one of them. In Section \ref{sec:hyper}, we identify the hyperparameter region that gives better decoding for cubic tensor codes, finding that it differs from the one for square tensor codes. In Section \ref{sec:perfeval}, we provide performance evaluation in terms of BLER, BER, and number of iterations to decoding for a collection of cubic product codes. These results indicate that a much larger palette of code constructions is possible with this new design. In Section \ref{sec:discussion}, we make closing remarks.

\section{Tensor code encoding}
\label{sec:tensor}
\begin{figure}
    \centerline{\input{3Dmatrix_u}}
    \caption{A cubic tensor code of dimension $(n^3,k^3)$, where each row, column and tube forms an $(n,k)$ codeword, for a code rate of $(k/n)^3$.}
    \label{fig:figure1}
\end{figure}
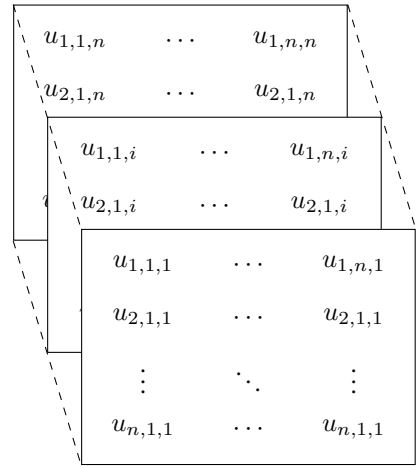
A binary tensor product code of $l$-dimensions is constructed using $l$ systematic binary linear component codes $\mathcal{C}_1$, $\mathcal{C}_2$,\ldots$\mathcal{C}_l$ \cite{elias_error-free_1954}. Each individual component code $\mathcal{C}_i$, $i \in \{1,\ldots,l\}$, is a $(n_i,k_i)$ code with rate $R_i = k_i/n_i$. The total number of information bits encoded by the code is $K = \prod_{i=1}^{l} k_i$. Encoding is achieved by first arranging the $K$ bits into an $l$-dimensional array, $k_1\times k_2 \times \ldots \times k_l$. From the $l$-dimensional array, each 1D slice of length $k_i$ is encoded with the corresponding component code $\mathcal{C}_i$. The resulting array becomes $k_1 \times k_2 \times \ldots \times k_l \times n_i$. After encoding all 1D slices, the final dimensions become $n_1$ $\times$ $n_2$ $\times$ \ldots $\times$ $n_l$ resulting in the final number of encoded bits being $N = \prod_{i=1}^{l} n_i$ for a code-rate of $R=K/N =\prod_{i=1}^{l} R_i$. Most practical constructions consider the 2D case and, as in Elias and Pyndiah \cite{elias_error-free_1954,pyndiah_1998}, encode using a single $(n,k)$ systematic code for both dimensions. With a single component code, the final code dimensions are $(n^l,k^l)$ giving a rate of $R=(k/n)^l$.

In the 2D case with a common component code, the rate of the tensor product code becomes $R$ = $k^2/n^2$.  In the 3D case, we call the dimensions rows, columns and tubes, where the latter describes the third dimension. Indexing the entries of the array with $(a,b,c)\in\{1,2,\ldots,n_1\}\times\{1,2,\ldots,n_2\}\times\{1,2,\ldots,n_3\}$, we denote the binary value of the array entry by $u_{a,b,c}$, and each row, column, and tube forms a codeword. For the case where all three component codes have dimensions $(n,k)$, Fig. \ref{fig:figure1} depicts the structure of the resulting tensor product code, which has rate $R$ = $(k/n)^3$.

Given a single systematic $(n,k)$ code, each component encoding requires $k(n-k)$ binary multiplications. For a 2D tensor code, one starts with $k \times k$ information bits. All $k$ rows can be encoded in parallel, resulting in $k\times n$ bits, followed by all $n$ columns to generate the final $n\times n$ encoding. In total, this requires $(k+n)k(n-k)$ binary multiplications.

\begin{figure}[h]
\begin{center}
\includegraphics[width=1\columnwidth]{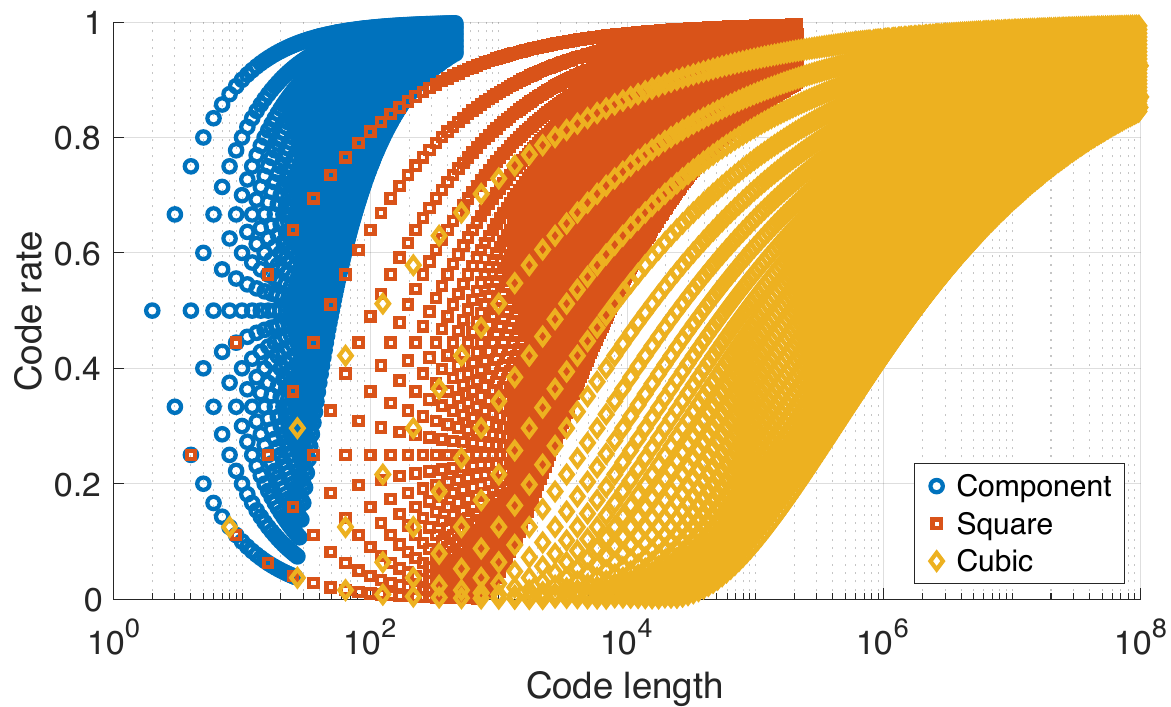}
\end{center}
\caption{For moderate redundancy component codes with $25$ or fewer redundant bits, available lengths and rates for component, square tensor product, and cubic tensor product codes with components of the same dimensions.
}
\label{fig:designspace}
\end{figure}

For a 3D code, one starts with $k \times k \times k$ information bits. One first encodes $k$ row component codes across $k$ tubes, for a total of $k^2$ component codes that can be encoded in parallel, resulting in a $k \times n \times k$ array. One can then encode $k\times n$ columns in parallel, resulting in a $n \times n \times k$ array. Finally, one can encode the $n \times n$ tubes in parallel, resulting in the final $n \times n \times n$ encoding. In total, this requires $(k^2+kn+n^2)k(n-k)$ binary multiplications.

For component codes with $n$ up to $464$ and $n-k\leq 25$, as could be readily decoded with SOGRAND \cite{yuan2023soft} and SO-GCD \cite{duffy2024SOGCD}, Fig. \ref{fig:designspace} illustrates the design space in terms of
rate and length that is available for square and cubic tensor product codes with $(n,k)$ components. For example, for a code of rate
$0.5$, the maximum length of component code alone is $50$, while for the square code it is $7,396$ and for the cubic code it
is $1,815,848$. In this way, longer, lower-rate, and, hence more powerful, codes can be constructed and then decoded with
large amounts of parallelism using one of the recent SISO decoders.

\begin{figure*}[h!]
    \centering
    \begin{minipage}{0.48\textwidth}
        \centering
        \includegraphics[width=\textwidth]{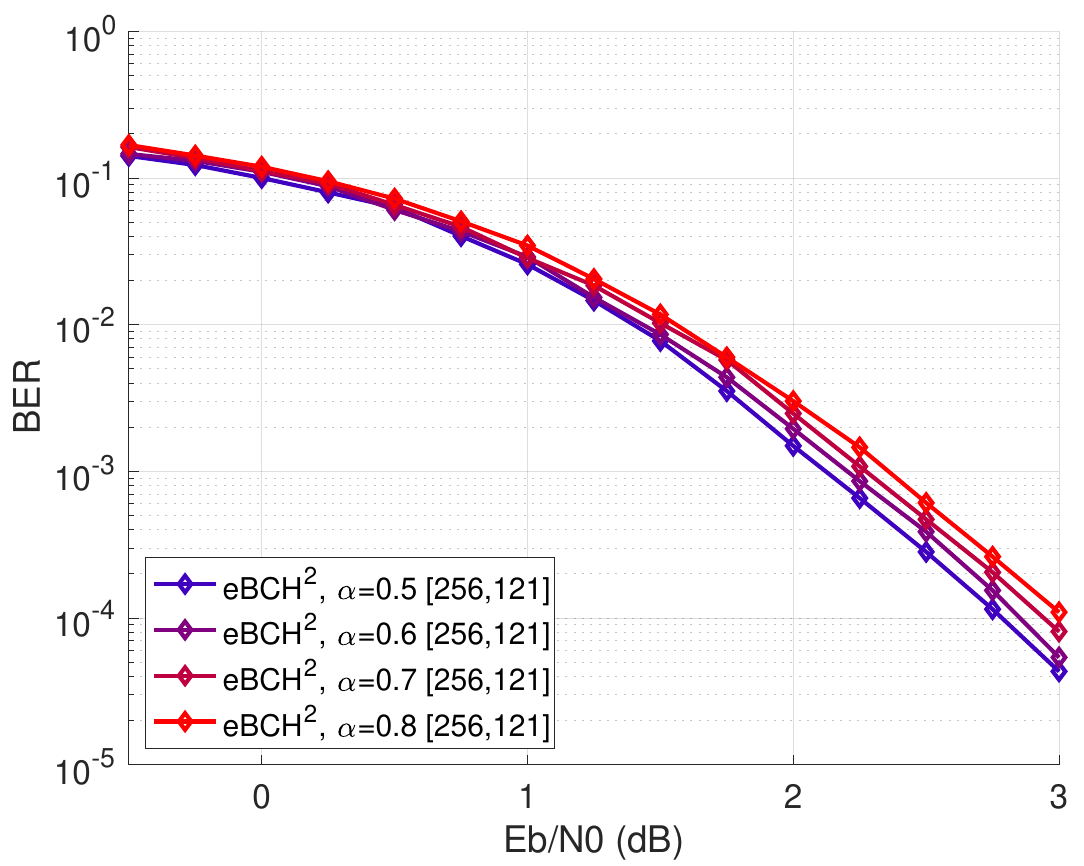}
        \caption{Performance of a [256,121] product code constructed using a [16,11] eBCH component code in an AWGN channel where iterative decoding is performed using SISO GRAND adapted for square product codes for different $\alpha$ values. Throughout decoding, a maximum of 20 iterations are considered with a list size L = 4 or an a posteriori list-BLER $<10^{-5}$ is reached.}

        \label{fig:2D-hyperparameter-sweep}
    \end{minipage}\hfill
    \begin{minipage}{0.48\textwidth}
        \centering
        \includegraphics[width=\textwidth]{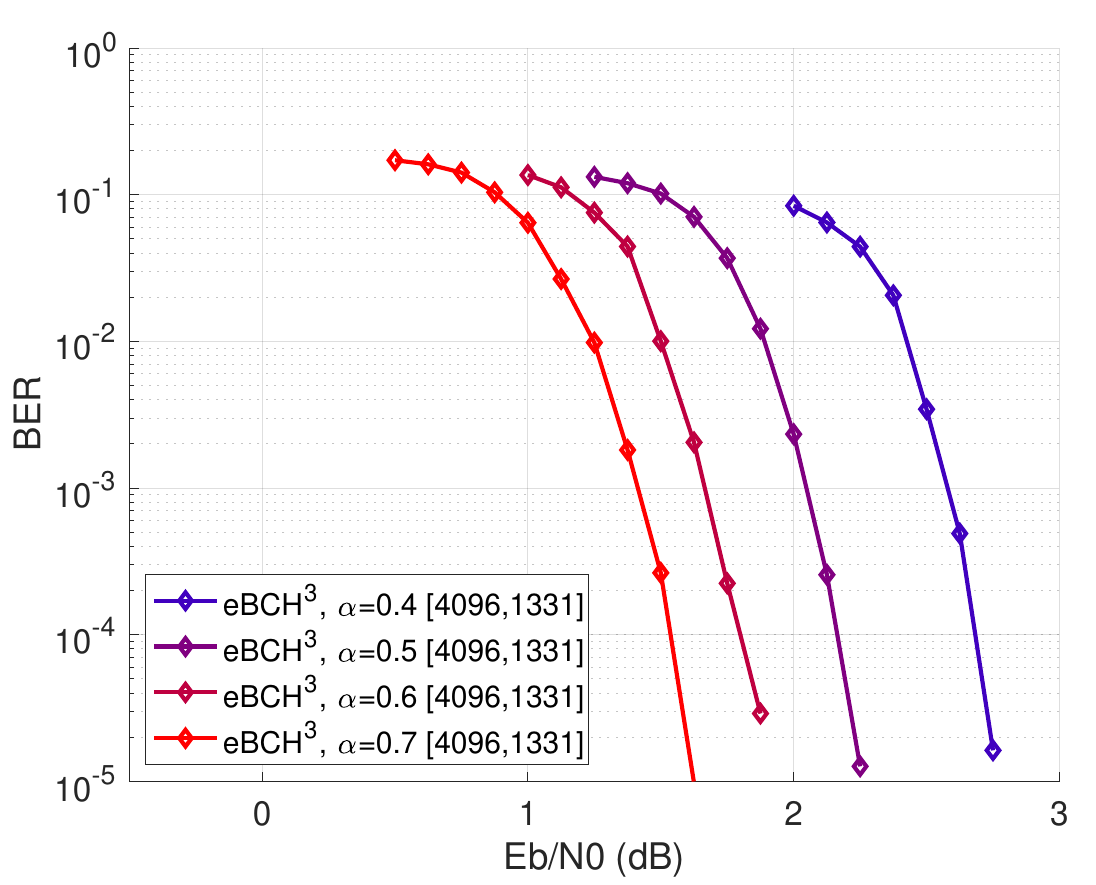}
        \caption{Performance of a [4096,1331] cubic tensor code constructed using a [16,11] eBCH component code in an AWGN channel where iterative decoding is performed using SISO GRAND adapted for cubic tensor codes for different $\alpha$ values. Throughout decoding, a maximum of 30 iterations are considered with a list size L = 4 or an a posteriori list-BLER $<10^{-5}$ is reached.}
        \label{fig:3D-hyperparameter-sweep}
    \end{minipage}
\end{figure*}
\begin{figure}[h]
\begin{center}
\includegraphics[width=1\columnwidth]{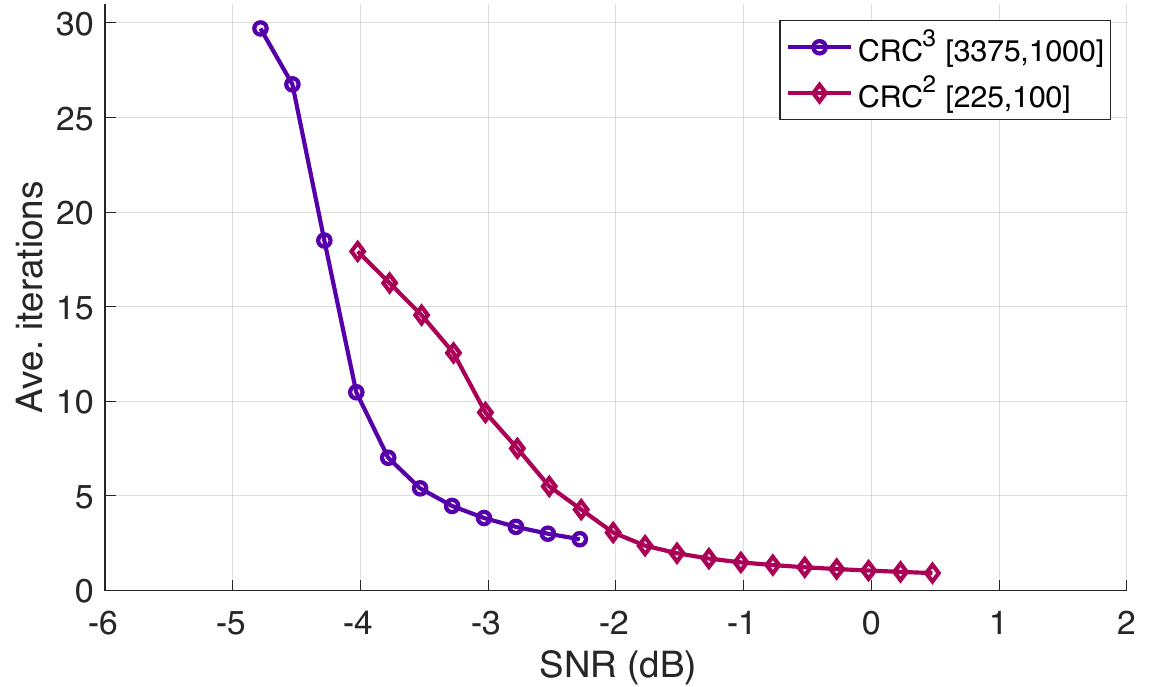}
\end{center}
\caption{Average number of iterations until a decoding is found for the square and cubic codes from Fig. \ref{fig:15_10_cubic}.
}
\label{fig:15_10_cubic_iter}
\end{figure}

\begin{figure}[h]
\begin{center}
\includegraphics[width=1\columnwidth]{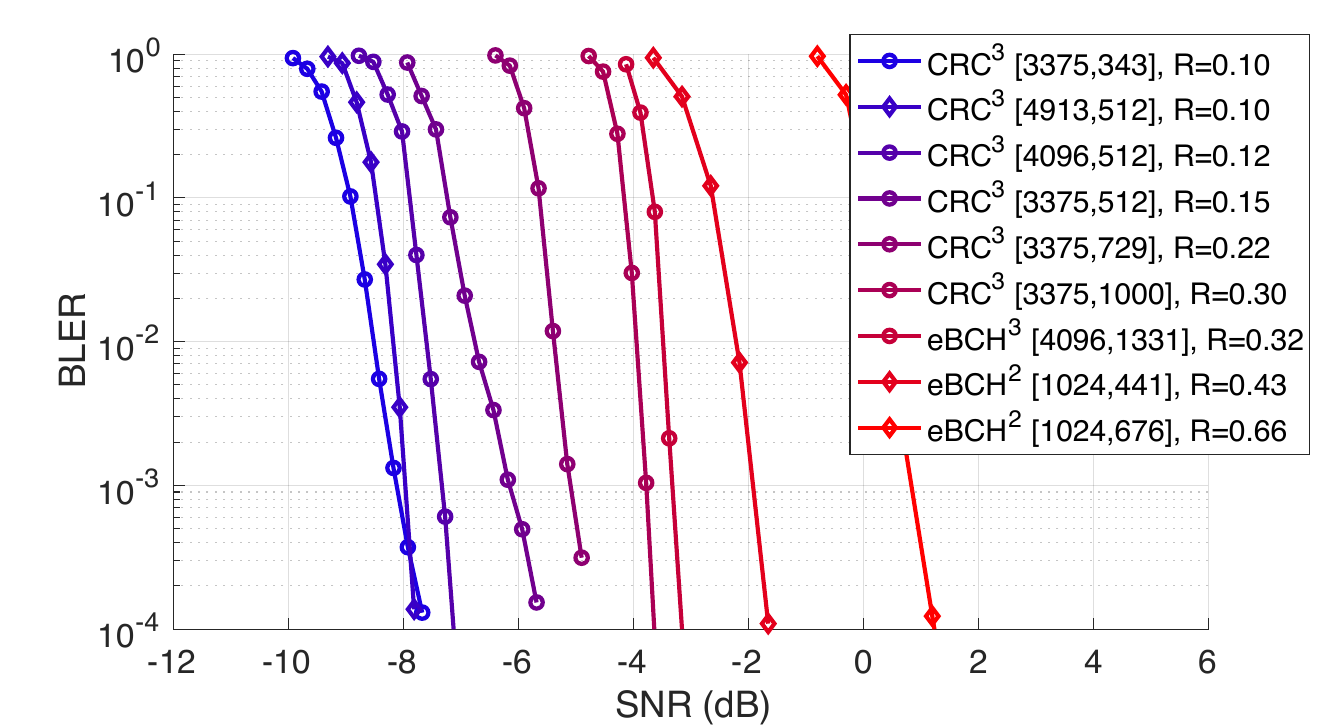}
\end{center}
\caption{Block error rate (BLER)  vs SNR (dB) for a collection of product and cubic codes. Component code decoding is iteratively performed with SOGRAND.} 
\label{fig:cubic_collection}
\end{figure}

\section{Turbo product code decoding}
\label{sec:TPC}
The iterative turbo product decoder assumes the existence of a SISO decoder for the component codes. The channel log-likelihood ratios (LLRs) $\boldsymbol{L}_{\text{Ch}_i}$ and the a-priori log-likelihood ratios $\boldsymbol{L}_{\text{A}_i}$ are the input of the SISO decoder for $i \in \{1,2,\ldots,n\}$ per row, column and tube. Per iteration, the SISO decoder outputs the a posteriori probabilities in the form of LLRs $\boldsymbol{L}_{\text{APP}_i}$ and the extrinsic LLRs $\boldsymbol{L}_{\text{E}}$. As with encoding, cubic tensor code decoding using a SISO decoder can be executed in parallel for the rows, columns, and tubes. 

Following Pyndiah \cite{pyndiah_1998}, but with one fewer hyperparameter, the algorithm for block turbo decoding is as follows:
\begin{itemize}
    \item[0] The channel LLRs are stored in a $n \times n \times n$ 3D array. During the first iteration, the a-priori LLRs of the received channel output is assumed to be equally likely. Hence, $\boldsymbol{L}_{\text{A}}$ is initialized to be a $\boldsymbol{0}$ of dimensions $n \times n \times n$. 
    \item[1] The SISO decoder processes each row of $\boldsymbol{L}$ = $\boldsymbol{L}_{\text{Ch}}$ + $\boldsymbol{L}_{\text{A}}$. The output APP LLRs and extrinsic LLRs are stored in their respective rows in $\boldsymbol{L}_{\text{APP}}$ and $\boldsymbol{L}_{\text{E}}$. If all the resulting rows, columns and tubes ($\mathbf{a}$,$i$,$\mathbf{c}$),($j$,$\mathbf{b}$,$\mathbf{c}$),($\mathbf{a}$,$\mathbf{b}$,$e$) for $ i,j,e \in \{1,2,\ldots,n\}$ yield a valid codeword $\mathbf{\hat{u}}_{\text{(a,b,c)}}$, the decoder returns the hard decision output $\mathbf{\hat{u}}_{\text{(a,b,c)}}$. If the iteration count (\textit{iter}) exceeds a set threshold (\textit{thres}), early termination is executed and a decoding failure is returned.  Otherwise, $\boldsymbol{L}_{\text{A}}$ $\leftarrow \alpha \boldsymbol{L}_{\text{E}}$, where $\alpha > 0$ is a pre-defined hyperparameter, and the decoder resumes to column updates.  
    \item[2] Every column of $\boldsymbol{L}$ is decoded updating the corresponding columns $\boldsymbol{L}_{\text{APP}}$ and $\boldsymbol{L}_{\text{E}}$. If the decision $\mathbf{\hat{u}}_{\text{(a,b,c)}}$ yields codewords per row, column and tube, the decoder returns $\mathbf{\hat{u}}_{\text{(a,b,c)}}$ as its hard decision. If \textit{iter} $>$ \textit{thres} an early termination is reached, as in 1. Otherwise, $\boldsymbol{L}_{\text{A}} \leftarrow \alpha \boldsymbol{L}_{\text{E}}$ to perform tube updates.
    \item[3]  Each tube of $\boldsymbol{L}$ is decoded and the corresponding tubes of $\boldsymbol{L}_{\text{APP}}$ and $\boldsymbol{L}_{\text{E}}$ are updated. If the hard decoder binary 3D array output $\mathbf{\hat{u}}_{\text{(a,b,c)}}$ is valid, decoding terminates by returning $\mathbf{\hat{u}}_{\text{(a,b,c)}}$. Similarly to 1, if \textit{iter} $>$ \textit{thres} an early termination is reached. Otherwise, $\boldsymbol{L}_{\text{A}}$ $\leftarrow \alpha {\boldsymbol{L}_{\text{E}}}$ and the next iteration starts (back to step 1).
\end{itemize}

\section{Hyperparameter}
\label{sec:hyper}

In Pyndiah's original turbo block decoder ~\cite{pyndiah_1998}, there are two hyperparameters: the weighing factor $\alpha$ and the reliability factor $\beta$. In the initial decoding iterations, the standard deviation of $\boldsymbol{L}_{\text{E}}$ is large, but it decreases in subsequent iterations. The weighting factor can change per half-iteration, corresponding to rows or columns, of decoding, and in ~\cite{pyndiah_1998} the values $\alpha = [0,0.2,0.3,0.5,0.7,0.9,1]$ were introduced as a scaling factor to reduce the effect of the extrinsic information in preliminary decoding iterations when the BER is large. The $\beta$ parameter represents the average reliability value of the soft decision bit output from the SISO decoder for which there is no competing codeword in the list decoding. In the original paper, both of the sets of parameters $\alpha$ and $\beta$ were experimentally determined.

The recently introduced SISO decoders, \cite{yuan2023soft,duffy2024SOGCD, yuan2024nearoptimal} produce accurate $\boldsymbol{L}_{\text{APP}}$ that dynamically weighs channel and decoding observations ~\cite{yuan2023soft,yuan2024nearoptimal}, circumventing the need for the reliability factor $\beta$ in the decoding process. In particular, here we will report on results using SOGRAND \cite{yuan2023soft} for which $\beta$ is not needed.

We performed a high-level hyperparameter search for the square product codes to confirm the reported choice of $\alpha= 0.5$ ~\cite{yuan2023soft,duffy2024SOGCD, yuan2024nearoptimal}. Fig. \ref{fig:2D-hyperparameter-sweep} presents a comparison of the different BER curves of a $(16^2$, $11^2) = (256,121)$ eBCH product code for different values of $\alpha$. Over the Eb/N0 (db) range, as previously reported, $\alpha$ in the region of $0.5$ yields the best performance relative to its counterparts in terms BER. 

Upon considering the cubic extension of the component code $(16^3,11^3) = (4096,1331)$, Fig. \ref{fig:3D-hyperparameter-sweep} illustrates the difference in BER performance with respect to different values of $\alpha$ in the region of $0.7$, finding it has a larger influence. At an Eb/N0 = 1.5 (db), a BER of approximately $2.6\times 10^{-4}$ is reached at $\alpha=0.7$ compared to a BER of $\approx 10^{-2}$ and $10^{-1}$ for $\alpha$ = 0.6 and 0.5 respectively. For the cubic code simulation results in the paper, $\alpha= 0.7$ is employed.

\section{Performance Evaluation}
\label{sec:perfeval}
For the simulation results, we consider the standard complex additive white Gaussian noise (AWGN) under binary phase shift keying (BPSK) modulation. The results in Fig.~\ref{fig:15_10_cubic} demonstrate that constructing square and cubic product codes using the same $(15,10)$ CRC component code results in longer, more powerful codes with decreasing rate. The component code has rate $2/3\approx0.67$, the square code is $4/9\approx0.44$ and the cubic code is $8/27\approx0.3$.

Consistent with the definition for LDPC codes and square product codes, decoding all rows, all columns or all tubes counts as a half iteration so that one full SISO decoding pass of a cubic code requires $1.5$ iterations. For the square and cubic codes reported in Fig. \ref{fig:15_10_cubic}, Fig. \ref{fig:15_10_cubic_iter} reports the average number of iterations until a decoding is found or the attempt abandoned, with the abandonment threshold for the square code being $20$ iterations and $30$ for the cubic code. It can be seen that the cubic code requires a small number of iterations and, indeed, fewer than the square code at a given SNR indicating that high-throughput, low-latency implementations would be possible in hardware.

\begin{figure}[h]
\begin{center}
\includegraphics[width=1\columnwidth]{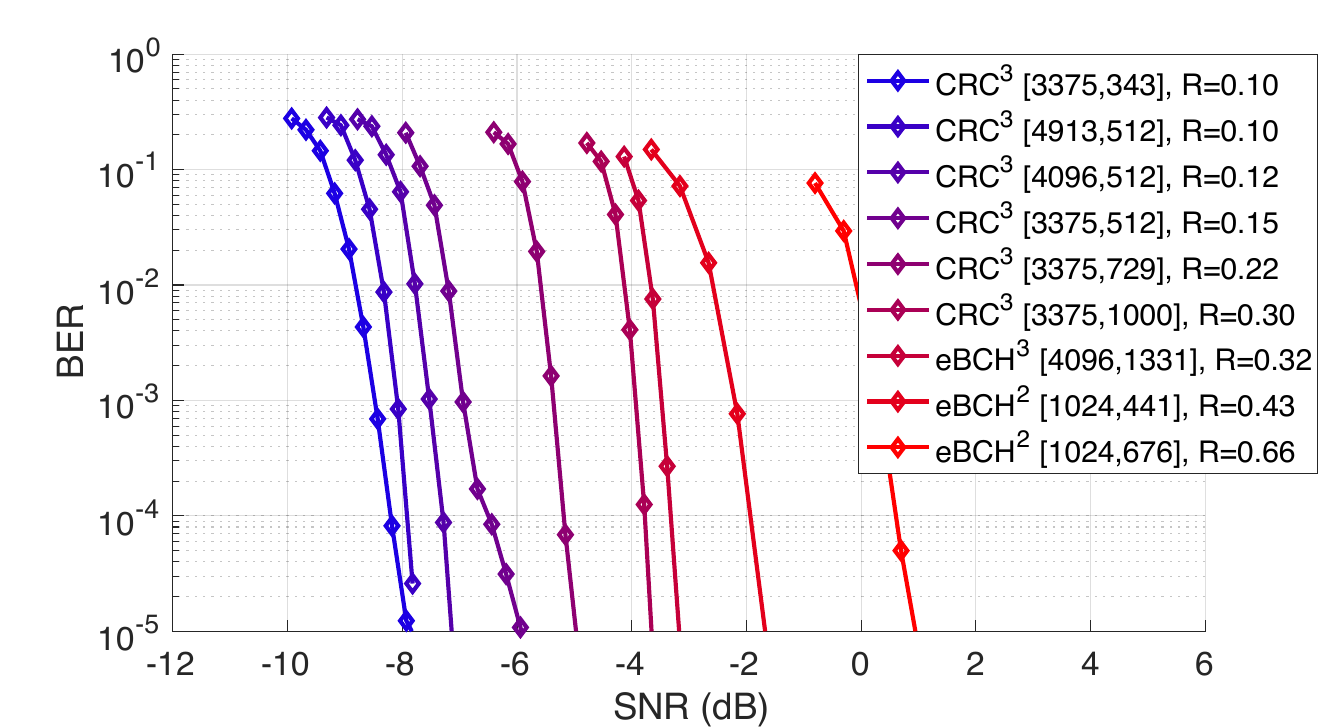}
\end{center}
\caption{Bit error rate (BER) vs SNR (dB) for a collection of product and cubic codes. Component code decoding is iteratively performed with SOGRAND.} 
\label{fig:cubic_collection_ber}
\end{figure}

\begin{figure}[h]
\begin{center}
\includegraphics[width=1\columnwidth]{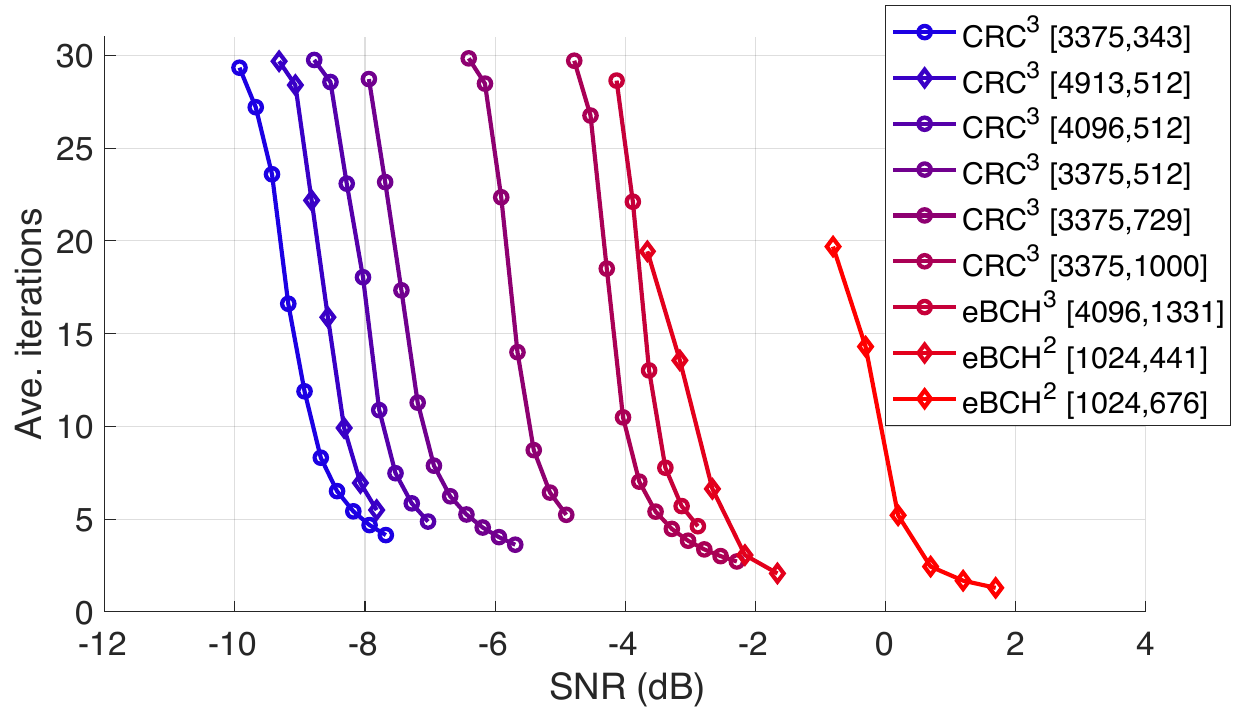}
\end{center}
\caption{Average number of iterations until decoding or abandonment vs SNR (dB) for the collection of product and cubic codes. Component code decoding is iteratively performed with SOGRAND.}
\label{fig:cubic_collection_iter}
\end{figure}

In order to assess the ability of cubic codes to provide reliable error correction at lower SNR, a wide range of cubic codes were constructed and decoded with SOGRAND \cite{yuan2023soft}. As SOGRAND can decode any component code, we availed of eBCH component codes as well as cyclic redundancy check codes that were identified by Koopman as they have been identified to have excellent error correction capabilities \cite{An21} when decoded with GRAND.

Fig. ~\ref{fig:cubic_collection} shows the performance of tensor product codes of different dimensions in terms of block error rate (BLER). The CRC cubic tensor product coded decoded iteratively using SOGRAND with $\alpha = 0.7$ are designed to give rates $\approx0.1$ -- $0.2$. The cubic CRC [3375,343] and [4913,512] have different length but essentially the same rate, $R\approx 0.10$, which are the lowest illustrated here. At around $8.1$ db and $8$ db, respectively, for which approximately $28$\% of demodulated bits are in error, the [3375,343] and [4913,512] CRC$^3$ codes exhibit a BLER of $10^{-3}$. Fig. \ref{fig:cubic_collection_ber} reports the post-decoding BER. At $\approx 8$dB, the two strongest codes take a demodulated BER where more than $1/4$ of the bits are in error to a post-decoding BER of less than $1$ in $1000$, demonstrating the error-correction capabilities of these constructions.

A proxy for decoding latency is the average number of decoding iterations until a decoding decision is made. Fig. ~\ref{fig:cubic_collection_iter} reports the average number of iterations before SOGRAND yields a decoding or abandonment for cubic and square product codes. Recall that the decoding of each of all rows, columns, and tubes is parallelizable, and each is regarded as a $1/2$ iteration. All of the CRC$^3$ and eBCH$^3$ codes yield a similar number of decoding iterations, on average, when compared to square product codes, even though the cubic codes here are of the order of $\approx2$ times longer than the square product codes. This desirable property exhibits flexibility and modularity of constructing a powerful, low rate code from components that can be tailored to system requirements while retaining a highly parallelizable decoding structure of the cubic codes using SOGRAND, SO-GCD or SO-SCL.

\section{Discussion}
\label{sec:discussion}
Traditional SI FEC applications, such as in wireless communications, typically require moderately powerful codes to compensate for physical layer errors and provide desired performance \cite{goldsmith2005wireless}. Highly challenged environments and some emerging applications, however, necessitate more powerful, low rate-codes. Such applications have been driving the construction of multi-edge type (MET) and cascade-structure LDPC codes, for example, that are been designed to provide rates between 10$^{-1}$ and 10$^{-2}$ ~\cite{mani2020error,richardson2008modern}. With the development of accurate SO decoders that can decode a broad range of component codes, \cite{yuan2023soft,yuan2024nearoptimal,duffy2024SOGCD}, here we explore an alternative construction: higher dimensional tensor codes. 

In this work, we demonstrate the performance of SISO GRAND decoding of low-rate, long cubic tensor codes with minimal latency in a highly parallelizable algorithm. The design space of tensor codes is large owing to SOGRAND and SO-GCD's ability to provide accurate SO for any moderate redundancy component code, while a large class of polar-like codes could be used with SO-SCL.

Prior work on square product codes suggests that $\alpha\approx0.5$ is a reasonable choice for most codes to provide good decoding parameters. Here, we find that, the single, static, fixed value for cubic tensor codes should be increased to $\approx0.7$. While it would be possible to further optimize decoding per cubic code by empirically determining an optimal $\alpha$, which may be dependent on the code, the SNR and indeed the iteration, for parsimony, a single $\alpha$ was selected for each evaluation.

Circuit and in-silicon implementations of ORBGRAND demonstrate that GRAND's code-book queries can be made in parallel, e.g. ~\cite{abbas2021orbgrand,condo2021fixed,Riaz23,Riaz24}. Coupled with leveraging the structure of tensor codes in the encoding and decoding process, low-latency can be attained even at higher dimensions with low-energy circuitry. The parallelizable decoding of the component codes per dimension may be incorporated by increasing the number of ORBGRAND circuits in a chip to result in low-latency decoding of long, powerful error correction codes. The flexibility of tensor product code construction in terms of component codes makes them readily amenable to desired lengths and rates without altering essential decoding architecture. 

\bibliographystyle{IEEEtran}
\bibliography{references}
\end{document}

%% file: 3Dmatrix_u.tex
\setbox0=\hbox{
        \begin{tikzpicture}[every node/.style={anchor=north east,fill=white,minimum width=1.4cm,minimum height=7mm}]
        \matrix (mA) [draw,matrix of math nodes]
        {
            u_{1,1,n}  &  \cdots & u_{1,n,n} \\
            u_{2,1,n} &  \cdots & u_{2,1,n} \\
            \vdots   &  \ddots & \vdots         \\
            u_{n,1,n}  &  \cdots & u_{n,1,n} \\
        };
        \matrix (mB) [draw,matrix of math nodes] at ($(mA.south west)+(4.9,1.65)$)
        {
            u_{1,1,i}  &  \cdots & u_{1,n,i} \\
            u_{2,1,i} &  \cdots & u_{2,1,i} \\
            \vdots   &  \ddots & \vdots         \\
            u_{n,1,i}  &  \cdots & u_{n,1,i} \\
        };
        \matrix (mC) [draw,matrix of math nodes] at ($(mB.south west)+(4.9,1.65)$)
        {
            u_{1,1,1}  &  \cdots & u_{1,n,1} \\
            u_{2,1,1} &  \cdots & u_{2,1,1} \\
            \vdots   &  \ddots & \vdots         \\
            u_{n,1,1}  &  \cdots & u_{n,1,1} \\
        };
        \draw[dashed](mA.north east)--(mC.north east);
        \draw[dashed](mA.north west)--(mC.north west);
        \draw[dashed](mA.south west)--(mC.south west);
        \end{tikzpicture}%
}

\kern10pt%
  \copy0